# The Effect of Multiple Scattering on the Polarization from Binary Star Envelopes.

## I. Self- and Externally Illuminated Disks

Short title: Polarization from Binary Star Envelopes. I.

Jennifer L. Hoffman[1,2], Barbara A. Whitney[3], and Kenneth H. Nordsieck[2]

*jhoffman@rice.edu, bwhitney@colorado.edu, khn@sal.wisc.edu*

---

[1] Department of Physics and Astronomy MS-108, Rice University, 6100 Main Street, Houston, TX  77005
[2] Department of Astronomy, University of Wisconsin-Madison, 475 N. Charter Street, Madison, WI  53706
[3] Space Science Institute, 3100 Marine Street, Suite A353, Boulder, CO  80303

ABSTRACT

We present the results of a Monte Carlo radiative transfer code that calculates the polarization produced by multiple Thomson scattering and variable absorptive opacity in a circumstellar disk around one component of a close detached binary system. We consider in detail the polarization variations over the binary cycle that result from the disk's illumination by the external star and by its own volume emission. We identify key features of these polarization phase curves and investigate their behavior as functions of optical depth, albedo, and inclination for geometrically thin and thick disks. The polarization due to disk self-illumination is sensitive to the internal optical characteristics of the disk, while the polarization arising from external illumination is mainly sensitive to the disk's geometrical thickness. With appropriate flux weighting, these results, combined with those for an internally illuminated disk, allow simulation of the polarization signature from an arbitrary binary-disk system.

Subject headings: circumstellar matter — methods: numerical — polarization — radiative transfer — scattering — binaries: general

1. INTRODUCTION

It is well known that circumstellar disks polarize starlight via electron or dust scattering. In the study of single stars with disks (Be stars, T Tauri stars), analysis of this polarization is used to extract information about the characteristics of the disks and the relation of each to its stellar system. But though circumstellar disks often appear in binary systems as well, the more complicated illumination geometries of these cases have limited the study of their polarization characteristics to the determination of system inclinations for the calculation of stellar masses and radii (e.g. Manset & Bastien 2000; Rudy & Kemp 1971). However, recent advances in numerical radiative transfer capabilities now allow us to treat binary systems in more detail and use polarization analysis to understand the properties of their circumstellar disks. Such efforts can yield important information regarding the structure and evolution of binary systems and allow comparisons between binary and single-star disks across the H-R diagram.

In their canonical theoretical study, Brown, McLean, & Emslie (1978, hereafter BME) derived analytical descriptions of the polarization arising from optically thin, corotating electron-scattering envelopes of arbitrary shape illuminated by point sources. Rudy & Kemp (1978) independently constructed very similar models. These results were extended by Brown et al. (1982) to eccentric orbits; by Simmons (1983) to arbitrary scattering mechanisms; by Fox & Brown (1991) to occultation effects; by Fox (1994) to finite illumination sources; and by Manset & Bastien (2000, 2001) to high optical depths and various grain sizes.

All these models used the simplifying assumption that the polarization arises from light that has been scattered only once. However, as shown by Wood et al. (1996a,b), multiple scattering often produces very different polarization results from those calculated using the single scattering and "single scattering plus attenuation" assumptions. In addition, the models above assumed that all illuminating sources lie within the scattering region, while in reality a disk often forms around only one component of a binary system. A full investigation of the polarization in binary-disk systems requires a treatment that takes into account multiple scattering effects and external as well as internal illumination sources. Monte Carlo radiative transfer codes are well suited for modeling such complex environments, as the numerical methods employed by these codes allow them to treat scattering regions that are optically thick, asymmetric, or nonuniform. Monte Carlo codes have proven to be of great use in analyzing the polarization arising from the circumstellar material surrounding single stars (e.g. Whitney, Kenyon, & Gomez 1997; Wood, Bjorkman, & Bjorkman 1997; Wood et al. 1996a,b; Whitney & Hartmann 1992, 1993). Daniel (1980) and Dolan (1984) began to explore the capabilities of Monte Carlo methods in the study of the polarization from binary-disk systems. Berger & Ménard (1997) presented Monte Carlo simulations of phase-averaged polarization in dusty shells around pre-main sequence binaries; Schultz (2000) did the same for the thin magnetic disks in low-mass X-ray binaries. Kurosawa & Hillier (2001) developed a sophisticated Monte Carlo code that can treat many complex scenarios. However, no systematic investigation exists of the effects of multiple scattering on the polarization arising from disks in binary systems. We here present such a study, using the results of a Monte Carlo code to investigate the polarization signatures resulting from a disk's illumination by its own volume emission and by the light from an orbiting companion. In a subsequent paper (Hoffman et al. 2003, hereafter Paper II), we invert the problem and show how analysis of the observed polarization of a binary system using the results of this paper can yield clues to the geometrical and optical characteristics of the circumstellar material within the system.

2. THE MONTE CARLO BINARY CODE

Our Monte Carlo binary star-disk code, called "DISK," is based on the "blobs" code described by Code & Whitney (1995); it is similar to the code presented by Wood et al. (1996a,b). Unpolarized virtual photons originate from an illumination source or sources, move through the scattering region with step sizes chosen from a probability distribution based on the optical depth of the region, scatter (and possibly become polarized) or become absorbed after each step, and are tabulated upon leaving the system. We simulate the effect of absorptive opacity via a variable scattering albedo ($a = 0$ means all interactions absorb photons, $a = 1$ means all interactions scatter photons). Wood et al. considered a star of either point or finite size illuminating a disk-shaped scattering region composed solely of electrons; we add to this geometry an external, finite star (Figure 1). The user specifies the radii of both stars, the distance between the stars, and the radius, opening angle, albedo, and equatorial optical depth of the disk. The addition of the external star breaks the longitudinal symmetry of the single star-disk system; we therefore bin outgoing photons by longitude ($\phi$) as well as latitude ($\theta$, equivalent to the inclination angle *I*). Our longitude bins are equally spaced; latitude bins have equal size in $\cos\theta$. Because the disk lies in the equatorial plane of the coordinate system, we can simulate the rotation of the binary (assuming circular orbits) by considering the photons in consecutive $\phi$ bins at a given $\theta$—in essence, by rotating the observer longitudinally around the coordinate system. Thus, a longitude $\phi$ (0–180°) can be converted into a phase (0–1) of the binary cycle.

Since the scattering disk depicted in Figure 1 is symmetric about the orbital plane of the binary and static in a corotating frame, we expect that when optically thin and illuminated by point sources, its polarization vector will describe a single ellipse in the *Q–U* plane, as predicted by BME. To compare our code results with BME's analytic formulation, we performed a test run in which we set the radii of the two stars to 0 and assigned the disk an opening angle $\alpha = 3°$, a small equatorial optical depth $\tau_{eq} = 0.1$, and an albedo $a = 0.95$ (slightly less than 1 to keep CPU times reasonable). The resulting *Q–U* curves for four representative inclination angles are shown in Figure 2; the model results are shown as points, while solid ellipses represent least-squares fits to the points. Figure 2 may be compared with BME's Figure 7, which depicts theoretical results for the case of an envelope symmetric about the orbital plane. Figure 3 compares the eccentricities of the ellipses fitted to our model results (*points*) with those predicted by BME (*dashed line*). Systematic errors in eccentricity introduced by the ellipse fitting routine have been subtracted from the data points in this figure (see Appendix for details). Our model reproduces the theoretical case very well. We note that the random errors generated by the fitting routine, shown as *y*-error bars in Figure 3, are much larger than those derived from the Stokes errors in our Monte Carlo model output.

We now turn to more complex scenarios. In our geometry (Figure 1), the disk is the only scattering region; it may be illuminated by three distinct light sources, giving rise to three different polarization signatures. Light may arise from the central star in what we call the "internally illuminated" case, from the external companion in the "externally illuminated" case, or from the disk itself in the "self-illuminated" case. Because the first scenario (internal illumination) has been modeled in detail by Wood et al. (1996a,b), we treat it only briefly, showing how our results compare with their previous work. We focus primarily on modeling the polarization created by self- and externally illuminated disks.

In the next section, we present the variation of polarization with the albedo and equatorial optical depth of the disk and the inclination angle of the system for two geometrical cases: a thin disk (opening angle $\alpha = 3°$) and a thick disk ($\alpha = 33°$), chosen to correspond to the thin and thick disks of the single-star case presented by Wood et al. We note that our disk is represented by a spherical wedge (Figure 1) with a constant electron density, while Wood et al. considered a spherical circumstellar region with $n_e$ obeying a power law in radius and $\sin \theta$. The geometrical configurations of the circumstellar material in the two studies are therefore slightly different. We will address the difference at appropriate junctures in the analysis that follows. Albedo is constant throughout the volume of our disk. In the self-illuminated case, photons are emitted uniformly and isotropically throughout the disk volume. To accomplish this, we randomly choose each photon's initial three position coordinates from the range of values spanned by the disk; similarly, we select each photon's initial direction of travel at random from the full $4\pi$ steradians. Our internal star has a radius of 0.1 $R_{disk}$; in the self- and externally illuminated cases, it does not emit photons, but still absorbs any whose paths intersect its volume. Our external star is 2 $R_{disk}$ away from the center of the disk and has a radius of 0.65 $R_{disk}$. We sort the outgoing photons into 35 latitude bins (spaced evenly in $\cos \theta$) and 72 longitude bins (spaced evenly in $\phi$); errors in each longitude-latitude bin are due to Poisson statistics and are calculated as described in Wood et al. (1996a). To obtain the results presented here, we ran $2 \times 10^8$ photons per model, with CPU times on a 450 MHz desktop PC ranging from 2 to 4 hours, depending on input parameters.

3. MODEL RESULTS

In presenting the polarization results of our code, we distinguish between photons that scattered only once in the circumstellar disk before exiting the system (singly scattered photons) and those that scattered twice or more (multiply scattered photons); we also consider the behavior of all scattered photons taken together. In the following figures, we compare the polarization characteristics of the singly scattered photons with those of all scattered photons.

Using subscripts to represent the number of scatters and lower-case $q$ to represent the ratio of Stokes parameters $Q/I$, we define

$$q_{\text{single}} = \frac{q_1 I_1}{I_0 + I_1} \tag{1}$$

and

$$q_{\text{all}} = \frac{q_1 I_1 + q_{2+} I_{2+}}{I_0 + I_1 + I_{2+}} \tag{2}$$

(similar equations hold for $u \equiv U/I$ ; see §3.2 for the case of $p \equiv \sqrt{q^2 + u^2}$ ). Including the unpolarized intensity $I_0$ in these quantities makes them more practical than the alternatives $q_{\text{single}} = q_1/I_1$ and $q_{\text{multiple}} = q_2/I_2$, which give the polarization level of each scattering category alone, but are not generally observed due to the difficulty of separating $I_1$ and $I_2$ from the total intensity. We have chosen to normalize the two quantities differently; with these definitions, $q_{\text{single}}$ may be compared with the results of binary star models that consider single scattering only, while $q_{\text{all}}$ may be compared with polarimetric observations. The reader should remember that because of this difference in normalization, one cannot assess the relative contributions of single and multiple scattering to $q_{\text{all}}$ from the information in the figures alone. One should instead refer to Tables 1–6, where we present for each of our model runs the numbers of photons escaping unscattered, singly scattered, and multiply scattered, as well as the number absorbed by the stars or disk. Note that the quantities in the tables are normalized to the number of photons originally emitted from the source. Although tabulated as percentages, they do not represent polarization, which we have not included in the tables because it varies substantially with viewing angle.

3.1. *Internal Illumination*

We first consider the case in which photons arise from the finite interior star; Tables 1 and 2 display the scattering statistics for the thin- and thick-disk cases. These statistics differ significantly from those of Wood et al. (1996b, their Tables 1 and 2) as a result of the differences in geometry between our model and those authors'. However, when the disk is thin (opening angle $\alpha = 3°$), our model produces polarization results nearly identical to those of Wood et al's Figures 2 and 3. For an equatorial optical depth $\tau_{eq} = 1$, polarization increases with increasing albedo; its magnitude peaks at ~1% at an inclination of ~75°. Table 1 shows that singly scattered photons account for most of this polarization signal. For $\tau_{eq} = 10$, most photons are still singly scattered, but multiple scattering becomes more important with increasing albedo; the magnitude of polarization is up to 3 times as large as for $\tau_{eq} = 1$, but shows the same behavior with inclination angle. As Wood et al. explained, this behavior results from the fact that increasing the equatorial optical depth and increasing the albedo both increase the number of scatterings the photons undergo. Photons that scatter multiple times within the disk midplane become highly polarized along the disk axis (in the positive $Q$ direction).

In the case of the thick disk, our results differ from those of Wood et al. (1996b), as we expect since the geometrical differences between our model disk and theirs (§2) are more prominent for a thicker disk. Our Figures 4 and 5 may be directly compared with Figures 5 and 6 of Wood et al. (Note that we use $q$ to denote the normalized Stokes parameter $Q/I$, where Wood et al. use capital Q; we represent the polarization arising from internal illumination with $q_{in}$.). For $\tau_{eq} = 1$ (Figure 4), our model displays qualitatively similar polarization behavior with albedo and inclination angle to that found by Wood et al. (their Figure 5). As in the thin-disk case, an increase in albedo produces more scatterings (Table 2) and thus more polarization. Since more photons interact with the thick disk, both the magnitude of polarization and the percentage of absorbed photons are systematically higher for all albedos than in the thin-disk case. In addition, the magnitude of polarization in our model is systematically higher than in the model of Wood et al. For $\tau_{eq} = 5$ (Figure 5), our model shows little difference from the $\tau_{eq} = 1$ case, except for a magnitude increase due to the increase in photon interactions. These results are very different from those of Wood et al (their Figure 6), which are characterized by large negative polarization values for singly scattered photons and a polarization minimum near $a = 0.4$ for multiply scattered photons. We attribute the differences between our results and those of Wood et al. to the difference in geometry between the two models. Scattering in the polar regions of Wood et al.'s geometry can give rise to negative polarization and thus decrease the overall polarization magnitude. Since no scattering occurs in the polar regions of our model, polar cancellation can normally only happen when photons reach the outer disk's higher latitudes after multiple scattering along the disk plane. This scenario requires quite high equatorial optical depths; at $\tau_{eq} = 10$ (not shown in Figure 5), we begin to see the effects of polar cancellation in our model, but still not to the extent found by Wood et al. For the relatively small optical depths shown in Figures 4 and 5, little if any cancellation occurs in our model. Slight negative polarization is produced only by singly scattered photons around $i = 70°$, when the favorable combination of inclination and optical depth allows these photons to scatter once in the high latitudes of the outer disk before reaching the observer.

3.2. *Self- and External Illumination*

We now turn to the main focus of this study: the polarization contribution of photons arising from the exterior star and from within the disk itself. We first present typical polarization curves for the self- and externally illuminated disks (both thin and thick) and identify quantities that define their polarization behavior over the course of the binary cycle. We defer discussion of the treatment of eclipses to Paper II, since the proper analysis of eclipse behavior must take into account the relative brightnesses of the system components. Here, we treat the self- and externally illuminated cases separately, and therefore examine their polarization behavior outside the eclipse phases (which we take to be 0.9–1.1 for eclipse of the external star and 0.4–0.6 for eclipse of the disk and internal star).

The quantity $q \equiv Q/I$ (expressed in percent) is a good choice to represent the polarization from the self-illuminated disk, since the Stokes parameter $Q$ represents the component of the polarization vector along the observer's $z$-axis (Figure 1), which aligns with the disk's axis when the system is inclined at 90° (edge-on) to the observer. With this geometry, all the polarization is in the $Q$ direction, and Stokes $U$ is zero. The percent polarization $q$ should, in principle, vary smoothly with inclination angle (Brown & McLean 1977). Since the disk's self-illumination is longitudinally symmetric, the resulting polarization is constant over the binary cycle (outside of eclipse), and we therefore denote it $q_{DC}$. For both the singly scattered photons and all photons together (Equations [1] and [2]), we take $q_{DC}$ to be a simple error-weighted average of $q$ over $\phi$ bins from phases 0.7–1.2 (Figure 6*a*).

In the case of the externally illuminated disk, $q$ is not a good index, since at lower inclinations, when the disk is oriented nearly face-on to the observer, the polarization will rotate between Stokes $Q$ and Stokes $U$ as the external star illuminates the disk from different $\phi$ directions. Instead, we use the total percent polarization $p \equiv \sqrt{q^2 + u^2} = \sqrt{Q^2 + U^2}/I$, also expressed in percent, to describe these results. In this case we have, instead of Equations (1) and (2),

$$p_{\text{single}} = \sqrt{q_{\text{single}}^2 + u_{\text{single}}^2} = \frac{I_1}{I_0 + I_1}\sqrt{q_1^2 + u_1^2} = \frac{p_1 I_1}{I_0 + I_1} \qquad (3)$$

and

$$p_{\text{all}} = \sqrt{q_{\text{all}}^2 + u_{\text{all}}^2} = \frac{1}{I_0 + I_1 + I_{2+}}\sqrt{(q_1 I_1 + q_{2+} I_{2+})^2 + (u_1 I_1 + u_{2+} I_{2+})^2} \qquad (4)$$

(which is *not* equal to $(p_1 I_1 + p_{2+} I_{2+})/(I_0 + I_1 + I_{2+})$ because of the cross-terms introduced under the square root). As discussed above, we have normalized these quantities differently because we expect $p_{\text{single}}$ to be useful in comparison with single-scattering models and $p_{\text{all}}$ to be comparable with observations. Figure 6*b* displays a typical total polarization curve from an externally illuminated disk; it shows peaks in $p$ near phases 0.3 and 0.7, when the external star is positioned so that the angle between it, the disk, and the observer is nearly 90°. For both singly scattered photons and all photons together, we characterize the size of these "bumps" with the error-weighted mean of $p$ at the "peak" phases (0.25–0.35 and 0.65–0.75; Figure 6*b*), and denote this value $p_{\text{bump}}$.

For each of the two geometries (thin and thick disk) described in §2, we ran models with equatorial optical depths $\tau_{\text{eq}} = 1, 3, 5, 7,$ and 10 and albedos $a = 0.1, 0.5,$ and 0.9. Figures 7 through 9 illustrate the variation of $q_{\text{DC}}$ and $p_{\text{bump}}$ with albedo and optical depth; we remind the reader that to determine the relative importance of photons scattered singly and multiply in each case, one should refer to Tables 3–6, where we present the scattering statistics for the self-illuminated thin and thick disks and externally illuminated thin and thick disks, respectively. We next discuss Figures 7–9 in turn.

3.2.1  *Polarization as a Function of Albedo*

Figure 7 shows the variation of $q_{DC}$ and $p_{bump}$ with albedo for low and high equatorial optical depths in the thin- and thick-disk cases. The absolute value of both polarization indices increases nearly linearly with increasing albedo, as photons become more likely to scatter and less likely to be absorbed in the disk. At small albedos, most photons are singly scattered, but as albedo increases, multiple scattering increases the absolute value of the polarization. This effect tends to be largest when the optical depth is highest, since in these cases each photon interacts many times. In almost all cases, the polarization is positive, indicating that most of the scattering occurs in the disk plane. However, $q_{DC}$ becomes slightly negative for a thick disk with high optical depth and an edge-on inclination, when scattering in the higher latitudes of the disk becomes important; see §3.2.2 for a more detailed discussion of optical depth effects. We discuss the variation of polarization with inclination angle in §3.2.3.

Figure 7 shows that $q_{DC}$ tends to be more sensitive to the optical depth and albedo of the disk than does $p_{bump}$. This is because $q_{DC}$ arises from scattering within the volume of the disk, while $p_{bump}$ is dominated by photons scattering off the disk edge. For the same reason, $p_{bump}$ is very sensitive to disk opening angle; an increase in opening angle increases the solid angle of the scattering region as seen from the external star, and thus increases the number of photons from the external star that intercept this region (note that the scale for $p_{bump}$ increases by a factor of 2–10, depending on optical depth, when the opening angle increases by a factor of 11).

Comparison of Figure 7*a* with Figure 7*b* shows that a thick disk produces less self-illuminated polarization ($q_{DC}$) and more externally illuminated polarization ($p_{bump}$) than does a thin disk with the same optical parameters. In a thick disk, more photons originate in or scatter into the polar regions than in a thin disk; this introduces more polar cancellation and thus decreases $q_{DC}$. The thick disk also presents a larger scattering surface to the external star than does the thin disk; this enhances $p_{bump}$.

It is interesting to compare the polarization of the self-illuminated disk ($q_{DC}$, thick lines) with that of the internally illuminated disk. For a thin disk (see Figure 2 of Wood et al. 1996b or our description in §3.1; also compare Tables 1 and 3), the two cases show qualitatively similar behavior as a function of albedo. The self-illuminated disk, however, generates about twice as much polarization as does the internally illuminated disk. We attribute the higher level of polarization in the self-illuminated case to the fact that these photons arise within the volume of the disk and are therefore more likely to interact with the scattering material, either being absorbed or becoming highly polarized before they escape the system. In the internally illuminated case, more photons escape without scattering. For a thick disk (Figures 4*a* and 5*a*; Tables 2 and 4), the two cases again behave similarly with albedo, but here the self-illuminated disk produces much less polarization than the internally illuminated disk (about half as much for $\tau_{eq} = 1$, but only a few percent as much for higher equatorial optical depths). This is because a photon originating within the thick disk is more likely to reach the high disk latitudes and become negatively polarized than is a photon arising from the internal star.

3.2.2 *Polarization as a Function of Optical Depth*

Figure 8 shows the variation of $q_{DC}$ and $p_{bump}$ with equatorial optical depth for low and high albedos in the thin- and thick-disk cases. We find that the self-illuminated polarization $q_{DC}$ varies widely with optical depth, and does so in different senses depending on whether the disk is thin or thick. In a thin disk, $q_{DC}$ increases with $\tau_{eq}$ up to intermediate values of $\tau_{eq}$, and then levels off or falls slightly. This behavior reverses for a thick disk, where $q_{DC}$ remains roughly constant or decreases with $\tau_{eq}$. Both trends arise from the same underlying phenomenon. As the midplane of either disk becomes optically thick, fewer photons can propagate within the disk plane. Those that are not absorbed scatter into the disk's higher latitudes, where they tend to acquire either a small positive or a negative polarization. For the thin disk, the effect is relatively small and only seen at high optical depths; it is similar to the behavior found by Wood et al. (1996a; their Figure 8). Because the thick disk is spatially larger than the thin disk, more photons reach its upper latitudes, and they do so at lower equatorial optical depths. This results in an overall decrease in polarization with $\tau_{eq}$. Interestingly, we do not see this behavior in the thick internally illuminated disk (compare Figures 4*a* and 5*a*) because photons arising from the interior star require very high optical depths in order to reach the higher latitudes of our model disk. Self-illumination of a uniform disk, then, can produce an effect similar in character, if not in magnitude, to that produced by internal illumination of a disk with low-density polar regions.

Though the behavior of the externally illuminated polarization, $p_{bump}$, mimics that of $q_{DC}$ in Figures 8*a* and 8*b*, it does so on a smaller absolute scale and with smaller relative variations. As noted in §3.2.1, $p_{bump}$ is less sensitive to the disk optical depth than is $q_{DC}$, because the photons producing this polarization arise from outside the disk and usually scatter only once (Tables 5 and 6) before reaching the observer.

### 3.2.3 *Polarization as a Function of Inclination Angle*

Figure 9 shows the variation of $q_{DC}$ and $p_{bump}$ with $\sin^2 i$ for two albedos and three equatorial optical depths in the thin- and thick-disk cases. We have chosen to plot $\sin^2 i$ on the *x*-axis for comparison with optically thin models, which predict that the polarization arising from an internally illuminated disk is proportional to $\sin^2 i$ (e.g. Brown & McLean 1977). Figure 9 shows that this proportionality holds for the thin self-illuminated disk, but only at lower inclination angles and low equatorial optical depths. Departures from direct proportionality stem from the effects of disk self-occultation at higher inclinations; here, the observer's line of sight intersects the thickest regions of the disk, where few scattered photons escape. The internally illuminated disk also shows this behavior (see Figures 2*b* and 3*b* of Wood et al. 1996b or our description in §3.1; see also Wood et al. 1996a). $p_{bump}$ shows a similar downturn at higher inclinations, but levels off as *i* approaches 0°; this suggests that the external star produces a larger polarization signal by illuminating the disk face than by illuminating the disk edge.

In the case of the thick disk, $q_{DC}$ loses its $\sin^2 i$ dependence at optical depths higher than 1, and becomes roughly constant or shallowly increasing, again turning down near 90°. This behavior results from the fact that the self-illuminated polarization from the edge of the thick disk includes both positive and negative components, so it is smaller than the polarization arising from the disk face, which it replaces as the inclination increases. (By contrast, the thin disk edge is highly positively polarized, giving rise to the steep increase in polarization with inclination seen in Figure 9*a*.) At high optical depths and inclinations, the self-occultation effect is extreme, and negatively polarized photons account for most of the polarization. $p_{bump}$ is nearly flat for all inclinations, indicating that the externally illuminated polarization from the disk edge is similar in magnitude to that from the disk face, which it replaces at high inclinations.

4. DISCUSSION AND CONCLUSIONS

We have presented results from Monte Carlo models of the polarization signatures of a wedge-shaped disk of constant density illuminated by an internal finite source, by an external finite source, and by its own uniform volume emission. Several complicating factors could alter the polarization behavior from that described here. If the external star fills its Roche lobe, its shape will depart substantially from spherical (see Paper II), and gravity brightening will cause nonuniform illumination over its surface. If the two stars are close enough together, the disk could also be distorted from axial symmetry by the gravitational influence of the external star. Gas streams, hot or cool spots on the stars or disk, jets, shells, and density variations within the disk can also give rise to polarization effects. Our intent in this paper has not been to investigate all these possible complications, but rather to establish the general behavior of relatively simple cases which can be used as starting points for more detailed analysis of the observed polarization variations of binary-disk systems. With this in mind, we now discuss our results in terms of this practical application. We note that for the results to be useful, the observer must know the direction of the intrinsic axis of the system under consideration, and must have removed any interstellar polarization contribution.

We find that the behavior of a self-illuminated disk with varying albedo and equatorial optical depth is, in general, qualitatively similar to that of a disk illuminated by an interior star (§3.1). The largest differences occur for a thick disk with high equatorial optical depth, when polar cancellation becomes important in the self-illuminated case. Quantitatively, a self-illuminated disk may produce either higher or lower polarization levels than an internally illuminated disk, depending on the combination of disk opening angle and optical properties; self-illumination appears to be an especially important source of polarization for geometrically thin disks, where $q_{DC}$ may reach 6%. We also find that neither the self-illuminated disk nor the internally illuminated disk produces large amounts of negative polarization, and that small amounts of negative polarization (a few tenths of a percent) occur only when the disk is both geometrically and optically thick. If one observes a large negative polarization roughly constant over the binary cycle, one can safely infer that it arises either from a thick envelope with optically thin polar regions such as that considered in Wood et al. (1996b) or from a polar jet or plume (Wood et al. 1996a), not from self- or internal illumination of a uniform disk.

External illumination of a circumstellar disk produces increases in polarization, or "bumps," near the binary phases 0.3 and 0.7; we have investigated the variation in magnitude of these bumps with disk properties and inclination angle. Because we have chosen to express this quantity in terms of $p$ instead of $q$ (see §3.2), its sign is always positive. In practice, $p_{bump}$ should be measured not from zero but from whatever constant (or "DC") polarization exists in the observed polarization curve. We find that the size of the polarization bumps produced by an externally illuminated disk is very sensitive to the opening angle of the disk, since this determines the height of the disk edge, from which most of the light scatters. If the disk is thin compared with the radius of the external star, the bump size is no larger than 0.65%, while a disk height comparable with the external star's radius leads to bumps as large as 1.5%. The observer must use caution in applying these results, however, since at low inclinations the polarization produced by the external star approaches a nonzero constant with phase (in the limiting face-on case, orbital variations appear only in position angle, not in magnitude). This results in a "DC" contribution from the external star (Figure 10), which can make the contributions to the polarization curve very difficult to disentangle. We have not investigated this polarization behavior, since it becomes significant only at inclination angles below ~55°, where the binary system is likely to be noneclipsing. Our results are best applied to eclipsing binary systems, where eclipse analysis allows the observer to determine the relative brightnesses of the components and therefore combine the polarization contributions in appropriate ratios (see below and Paper II).

Since in practice the quantities $p_{\text{bump}}$ and $q_{\text{DC}}$ will exist in the same observational phase curve, it is useful to know how they compare with one another. The ratio $p_{\text{bump}}/q_{\text{DC}}$ gives a clue to the shape of the composite polarization curve of a binary system (in the absence of a contribution from an internal source, which we address below). We must keep in mind, however, that when we combine polarization results from two or more sources, we must specify their relative brightnesses as seen by the observer. This is because in practice we do not know how much of the total observed flux arises from each source, and therefore cannot measure $p_{\text{ext}} \equiv P_{\text{ext}}/I_{\text{ext}}$ or $q_{\text{disk}} = Q_{\text{disk}}/I_{\text{disk}}$; rather, the observed percent polarization we attribute to each source reflects that source's polarized flux divided by the *total* flux from the system. Specifying relative brightnesses allows us to make an appropriate comparison of the percent polarization arising from each source alone using the relation $p_{\text{ext}}/q_{\text{disk}} = (P_{\text{ext}}/Q_{\text{disk}}) \cdot (I_{\text{disk}}/I_{\text{ext}})$. Figure 11 shows the variation of the ratio $p_{\text{bump}}/q_{\text{DC}}$ with $\sin^2 i$ for two albedos and three equatorial optical depths in the thin- and thick-disk cases, assuming that the disk and external star are equally bright ($I_{disk} = I_{ext}$). In the thick-disk case, $q_{DC}$ is often small, and approaches zero for the highest inclinations, causing the ratio to be ill-behaved for these cases. We have therefore truncated the plots in Figure 11b at $\sin^2 i = 0.85$. The magnitude of $p_{bump}/q_{DC}$ behaves similarly to $p_{bump}$ (see §3.2.1): it is roughly proportional to the disk opening angle and is not sensitive to equatorial optical depth or to albedo. Typically, the ratio is a maximum at low inclinations, and decreases smoothly with increasing $i$. The contribution of multiple scattering causes $q_{DC}$ to increase more than $p_{bump}$, which results in a decrease in the $p_{bump}/q_{DC}$ ratio.

Assuming that the disk is unlikely to be intrinsically brighter than the external star, the results in Figure 11 may be viewed as upper limits: the brighter the star is with respect to the disk, the smaller $p_{bump}/q_{DC}$ will be. For the thin disk and equal brightnesses, $p_{bump}$ is much smaller than $q_{DC}$, and may therefore be difficult to observe; it will be more apparent in the polarization curve if the disk is thick and reasonably bright. The very appearance of bumps at phases 0.3 and 0.7 in the polarization curve of a binary star may thus provide evidence for a thick disk in the system. In fact, since the ratios for the thin and thick cases do not overlap and are insensitive to the disk's optical properties, measurement of $p_{bump}/q_{DC}$ (for an assumed brightness ratio and in the absence of other illumination sources) could lead directly to an estimate of both disk opening angle and inclination.

In practice, however, the contribution of an interior star to the phase-constant polarization cannot normally be ignored, and this makes the situation much more complex. Figure 12 displays the variation of the ratio $q_{DC}/q_{in}$ with $\sin^2 i$ for the same parameters as in Figure 11, with the assumption that the interior star and disk are equally bright. This ratio decreases with disk opening angle, but like $p_{bump}/q_{DC}$ is relatively insensitive to the disk's optical properties. $q_{in}$ may be larger or smaller than $q_{DC}$, depending on the opening angle and the relative brightnesses of the disk and interior star; it will be especially important if the disk is thick or very bright. In the case of a passive (nonradiating) disk, the ratio $p_{bump}/q_{in}$ is relevant; for the assumption of equal brightnesses, this ratio lies between 0 and 1 and decreases smoothly with $\sin^2 i$ (with occasional deviations when $q_{in}$ is near 0). In the interest of brevity, we do not display $p_{bump}/q_{in}$, but it can easily be calculated by multiplying the quantities in Figures 11 and 12 for a given set of disk parameters.

It is clear that using the results of this study in the analysis of the observed polarization from a binary-disk system requires the knowledge (or assumption) of several other system properties. We close with a brief outline of the steps required.

1. Remove the interstellar polarization contribution, if any.

2. Estimate the orientation of the intrinsic axis of the system, either from light-curve analysis or by inspection of the behavior of the position angle of the polarization (we used this latter method in Hoffman et al. [1998] to determine the orientation of the β Lyr system).

3. Rotate all polarization measurements to this axis so that $q_{DC}$ (now equivalent to "$p_{DC}$"; note that this quantity may include $q_{in}$) and $p_{bump}$ may be directly compared.

4. Measure $q_{DC}$ ($p_{DC}$) starting at zero polarization; measure $p_{bump}$ starting at the $q_{DC}$ ($p_{DC}$) level.

5. Estimate the relative brightnesses of the illuminators in the system (see Paper II for one method of doing this) and scale the results presented here as appropriate.

6. Compare the scaled model results with the observed quantities to constrain the properties of the disk.

In Paper II, we illustrate this method by applying it to the eclipsing binary system β Lyr, for which we published intrinsic polarization curves in Hoffman et al. (1998).

## ACKNOWLEDGMENTS

We extend many thanks to our anonymous referee for numerous thoughtful suggestions; JLH also thanks Jon Fowler for editing services and helpful discussions of ellipse fitting and its associated errors. We are heavily indebted to Mumit Khan of the University of Wisconsin's Engineering Department for his patient and invaluable assistance with Linux Fortran.

## APPENDIX
## LEAST-SQUARES ELLIPSE FITTING

We used the IDL least-squares fitting routine MPFITELLIPSE[4] to calculate the eccentricities of the $Q$–$U$ ellipses resulting from our BME comparison models described in §2. We found that these calculated eccentricities departed significantly from those prescribed by the BME formalism, especially at lower inclination angles. This is the same effect documented by Aspin, Simmons, & Brown (1981) and Simmons, Aspin, & Brown (1982), who showed that orbital inclinations inferred from simple least-squares fitting of observational polarimetric data are biased toward high inclinations (which correspond to high eccentricities; see Figure 3) by even a small amount of noise. In our particular case, the bias arises from the fact that a least-squares fitting routine that assumes the fit is an ellipse cannot return an eccentricity less than 0; the size of the bias increases with increasing data noise.

We investigated the extent of this bias with a procedure similar to that of Wolinski & Dolan (1994). For each of the five inclinations plotted in Figure 3, we calculated the "true" theoretical eccentricity of the *Q*–*U* ellipse from the equation $e = \sin^2 i/(2-\sin^2 i)$ (BME). We used the IDL random number generators to create pseudo-data points that mimicked our model output: 24 points evenly spaced in phase angle around an ellipse with the "true" eccentricity. We then introduced statistical noise by perturbing each point in *Q* and *U* by random amounts chosen from a Gaussian distribution with a mean of zero and a variance corresponding to the size of the error bars in our model data (8-9%). Finally, we applied MPFITELLIPSE to the pseudo-data. After repeating this procedure 100 times, we took the mean of the calculated eccentricities for each inclination to be the systematic error introduced by the fitting routine, and subtracted this value from the appropriate data point to produce Figure 3. The *y*-error bars in the figure represent the standard deviation of the 100 calculated eccentricities for each point.

---

[4] MPFITELLIPSE is a public-domain IDL program written by C. Markwardt; it is available from his web page, http://cow.physics.wisc.edu/~craigm/idl/fitting.html.

REFERENCES

Aspin, C., Simmons, J.F.L., & Brown, J.C. 1981, MNRAS, 194, 283

Berger, J.-P., & Ménard, F. 1997, IAUS, 182, 201

Brown, J.C., Aspin, C., Simmons, J.F.L., & McLean, I.S. 1982, MNRAS, 198, 787

Brown, J.C., & McLean, I.S. 1977, A&A, 57, 141

Brown, J.C., McLean, I.S., & Emslie, A.G. 1978, A&A, 68, 415 (BME)

Code, A.D. & Whitney, B.A. 1995, ApJ, 441, 400

Dolan, J.F. 1984, A&A, 138, 1

Fox, G.K. 1994, ApJ, 432, 262

Fox, G.K., & Brown, J.C. 1991, ApJ, 379, 663

Hoffman, J.L., Nordsieck, K.H., & Fox, G.K. 1998, AJ, 115, 1576

Hoffman, J.L., Nordsieck, K.H., & Whitney, B.A. 2003, in preparation (Paper II)

Kurosawa, R., & Hillier, D.J. 2001, A&A, 379, 336

Manset, N., & Bastien, P. 2000, AJ, 120, 413

———. 2001, AJ, 122, 2692

Rudy, R.J., & Kemp, J.C. 1978, ApJ, 221, 200

Schultz, J. 2000, A&A, 364, 587

Simmons, J.F.L. 1983, MNRAS, 205, 153

Simmons, J.F.L., Aspin, C., & Brown, J.C. 1982, MNRAS, 198, 45

Whitney, B.A., & Hartmann, L. 1992, ApJ, 395, 529

———. 1993, ApJ, 402, 605

Whitney, B.A., Kenyon, S.J., & Gomez, M. 1997, ApJ, 485, 703

Wolinski, K.G., & Dolan, J.F. 1994, MNRAS, 267, 5

Wood, K., Bjorkman, J.E., Whitney, B., & Code, A. 1996a, ApJ, 461, 828

———. 1996b, ApJ, 461, 847

Wood, K., Bjorkman, K.S., & Bjorkman, J.E. 1997, ApJ, 477, 926

TABLE 1
PERCENTAGES OF UNSCATTERED, SCATTERED, AND ABSORBED PHOTONS
FOR THE INTERNALLY ILLUMINATED THIN DISK

| Albedo | Equatorial Optical Depth | Unscattered Photons (%) | Singly Scattered Photons (%) | Multiply Scattered Photons (%) | Absorbed Photons (%) |
|---|---|---|---|---|---|
| 0.1…….. | 1  | 93.75 | 0.33 | 0.00 | 5.92 |
|         | 3  | 89.96 | 0.59 | 0.01 | 9.44 |
|         | 5  | 87.93 | 0.68 | 0.02 | 11.37 |
|         | 7  | 86.58 | 0.72 | 0.02 | 12.68 |
|         | 10 | 85.14 | 0.75 | 0.03 | 14.09 |
| 0.5…….. | 1  | 93.75 | 1.64 | 0.09 | 4.52 |
|         | 3  | 89.96 | 2.93 | 0.36 | 6.75 |
|         | 5  | 87.93 | 3.39 | 0.56 | 8.12 |
|         | 7  | 86.58 | 3.61 | 0.70 | 9.12 |
|         | 10 | 85.13 | 3.75 | 0.85 | 10.26 |
| 0.9…….. | 1  | 93.75 | 2.95 | 0.31 | 2.99 |
|         | 3  | 89.96 | 5.27 | 1.32 | 3.44 |
|         | 5  | 87.93 | 6.11 | 2.15 | 3.81 |
|         | 7  | 86.58 | 6.49 | 2.81 | 4.13 |
|         | 10 | 85.14 | 6.74 | 3.58 | 4.54 |

NOTE.—Percentages in this and all subsequent tables refer to the numbers of unscattered, scattered, and absorbed photons divided by the total number of photons emitted from the central star. These percentages should not be confused with polarization.

TABLE 2
PERCENTAGES OF UNSCATTERED, SCATTERED, AND ABSORBED PHOTONS
FOR THE INTERNALLY ILLUMINATED THICK DISK

| Albedo | Equatorial Optical Depth | Unscattered Photons (%) | Singly Scattered Photons (%) | Multiply Scattered Photons (%) | Absorbed Photons (%) |
|---|---|---|---|---|---|
| 0.1…….. | 1 | 63.57 | 1.90 | 0.08 | 34.45 |
|  | 3 | 43.84 | 1.44 | 0.10 | 54.61 |
|  | 5 | 39.39 | 1.10 | 0.08 | 59.43 |
|  | 7 | 37.67 | 0.94 | 0.07 | 61.32 |
|  | 10 | 36.24 | 0.81 | 0.06 | 62.89 |
| 0.5…….. | 1 | 63.57 | 9.50 | 2.53 | 24.40 |
|  | 3 | 43.83 | 7.23 | 3.79 | 45.15 |
|  | 5 | 39.38 | 5.49 | 3.07 | 52.05 |
|  | 7 | 37.66 | 4.71 | 2.63 | 55.00 |
|  | 10 | 36.23 | 4.05 | 2.30 | 57.41 |
| 0.9…….. | 1 | 63.56 | 17.11 | 10.45 | 8.88 |
|  | 3 | 43.84 | 13.02 | 22.93 | 20.21 |
|  | 5 | 39.38 | 9.89 | 22.51 | 28.22 |
|  | 7 | 37.66 | 8.47 | 20.37 | 33.50 |
|  | 10 | 36.23 | 7.29 | 18.07 | 38.41 |

TABLE 3
PERCENTAGES OF UNSCATTERED, SCATTERED, AND ABSORBED PHOTONS
FOR THE SELF-ILLUMINATED THIN DISK

| Albedo | Equatorial Optical Depth | Unscattered Photons (%) | Singly Scattered Photons (%) | Multiply Scattered Photons (%) | Absorbed Photons (%) |
|---|---|---|---|---|---|
| 0.1…….. | 1  | 87.31 | 0.81  | 0.01  | 11.87 |
|         | 3  | 75.72 | 1.54  | 0.04  | 22.70 |
|         | 5  | 67.79 | 1.82  | 0.07  | 30.33 |
|         | 7  | 61.78 | 1.91  | 0.09  | 36.22 |
|         | 10 | 54.97 | 1.90  | 0.11  | 43.03 |
| 0.5…….. | 1  | 87.31 | 4.06  | 0.26  | 8.37  |
|         | 3  | 75.72 | 7.70  | 1.24  | 15.34 |
|         | 5  | 67.79 | 9.07  | 2.15  | 20.98 |
|         | 7  | 61.79 | 9.54  | 2.87  | 25.81 |
|         | 10 | 54.97 | 9.48  | 3.60  | 31.95 |
| 0.9…….. | 1  | 87.31 | 7.30  | 0.88  | 4.50  |
|         | 3  | 75.72 | 13.85 | 4.64  | 5.79  |
|         | 5  | 67.79 | 16.33 | 8.70  | 7.17  |
|         | 7  | 61.78 | 17.17 | 12.43 | 8.62  |
|         | 10 | 54.97 | 17.06 | 17.08 | 10.88 |

TABLE 4
PERCENTAGES OF UNSCATTERED, SCATTERED, AND ABSORBED PHOTONS
FOR THE SELF-ILLUMINATED THICK DISK

| Albedo | Equatorial Optical Depth | Unscattered Photons (%) | Singly Scattered Photons (%) | Multiply Scattered Photons (%) | Absorbed Photons (%) |
|---|---|---|---|---|---|
| 0.1…….. | 1  | 60.25 | 2.16  | 0.09  | 37.50 |
|         | 3  | 32.72 | 1.89  | 0.14  | 65.25 |
|         | 5  | 22.33 | 1.38  | 0.12  | 76.18 |
|         | 7  | 17.09 | 1.05  | 0.09  | 81.76 |
|         | 10 | 12.89 | 0.76  | 0.07  | 86.27 |
| 0.5…….. | 1  | 60.24 | 10.80 | 2.70  | 26.26 |
|         | 3  | 32.72 | 9.45  | 4.98  | 52.85 |
|         | 5  | 22.33 | 6.91  | 4.46  | 66.30 |
|         | 7  | 17.09 | 5.27  | 3.71  | 73.93 |
|         | 10 | 12.89 | 3.83  | 2.84  | 80.44 |
| 0.9…….. | 1  | 60.24 | 19.44 | 10.99 | 9.32  |
|         | 3  | 32.72 | 17.00 | 28.95 | 21.32 |
|         | 5  | 22.33 | 12.43 | 32.71 | 32.52 |
|         | 7  | 17.09 | 9.49  | 31.35 | 42.07 |
|         | 10 | 12.89 | 6.89  | 27.13 | 53.10 |

TABLE 5
PERCENTAGES OF UNSCATTERED, SCATTERED, AND ABSORBED PHOTONS
FOR THE EXTERNALLY ILLUMINATED THIN DISK

| Albedo | Equatorial Optical Depth | Unscattered Photons (%) | Singly Scattered Photons (%) | Multiply Scattered Photons (%) | Absorbed Photons (%) |
|---|---|---|---|---|---|
| 0.1…….. | 1 | 99.51 | 0.04 | 0.00 | 0.45 |
| | 3 | 99.03 | 0.06 | 0.00 | 0.90 |
| | 5 | 98.78 | 0.07 | 0.00 | 1.15 |
| | 7 | 98.62 | 0.07 | 0.00 | 1.31 |
| | 10 | 98.49 | 0.06 | 0.00 | 1.44 |
| 0.5…….. | 1 | 99.51 | 0.18 | 0.01 | 0.30 |
| | 3 | 99.03 | 0.31 | 0.05 | 0.61 |
| | 5 | 98.78 | 0.34 | 0.08 | 0.80 |
| | 7 | 98.62 | 0.34 | 0.10 | 0.94 |
| | 10 | 98.49 | 0.32 | 0.11 | 1.08 |
| 0.9…….. | 1 | 99.51 | 0.32 | 0.04 | 0.13 |
| | 3 | 99.03 | 0.56 | 0.19 | 0.22 |
| | 5 | 98.78 | 0.62 | 0.32 | 0.29 |
| | 7 | 98.63 | 0.61 | 0.42 | 0.34 |
| | 10 | 98.49 | 0.58 | 0.52 | 0.41 |

TABLE 6
PERCENTAGES OF UNSCATTERED, SCATTERED, AND ABSORBED PHOTONS
FOR THE EXTERNALLY ILLUMINATED THICK DISK

| Albedo | Equatorial Optical Depth | Unscattered Photons (%) | Singly Scattered Photons (%) | Multiply Scattered Photons (%) | Absorbed Photons (%) |
|---|---|---|---|---|---|
| 0.1…….. | 1  | 96.69 | 0.19 | 0.01 | 3.12 |
|         | 3  | 95.04 | 0.16 | 0.01 | 4.79 |
|         | 5  | 94.57 | 0.13 | 0.01 | 5.29 |
|         | 7  | 94.36 | 0.11 | 0.01 | 5.52 |
|         | 10 | 94.21 | 0.10 | 0.01 | 5.69 |
| 0.5…….. | 1  | 96.69 | 0.93 | 0.22 | 2.16 |
|         | 3  | 95.03 | 0.79 | 0.34 | 3.84 |
|         | 5  | 94.58 | 0.65 | 0.31 | 4.46 |
|         | 7  | 94.36 | 0.57 | 0.28 | 4.78 |
|         | 10 | 94.20 | 0.51 | 0.25 | 5.04 |
| 0.9…….. | 1  | 96.69 | 1.68 | 0.88 | 0.75 |
|         | 3  | 95.04 | 1.42 | 1.86 | 1.67 |
|         | 5  | 94.56 | 1.17 | 1.98 | 2.29 |
|         | 7  | 94.36 | 1.03 | 1.91 | 2.70 |
|         | 10 | 94.21 | 0.91 | 1.78 | 3.10 |

Fig. 1—Sketch of the star-disk geometry used for our Monte Carlo models. The disk, shown in cross-section, is a spherical wedge of constant density with opening angle $\alpha = 3°$ or $33°$; the external star is 2 $R_{disk}$ away and spherical with a radius 0.65 $R_{disk}$. The inner star is spherical with a radius of 0.1 $R_{disk}$.

Fig. 2—*Q*–*U* diagrams of the polarization from a Monte Carlo model simulating a geometrically and optically thin disk illuminated by two point sources and viewed from various inclination angles. The heavy lines and labels refer to best-fit ellipses. Error bars at lower left show representative Stokes errors for each case.

Fig. 3—(*a*) Comparison of the eccentricities of the best-fit ellipses to the *Q*–*U* curves of our Monte Carlo test case (see Fig. 2; *y-axis*) with the theoretical results of Brown et al. (1978) *(x-axis)*. The dashed line represents exact agreement between the two. Error bars in the *x*-direction arise from the finite width of our inclination angle bins; error bars in the *y*-direction arise from random errors in the ellipse fitting procedure. Systematic errors introduced by the fitting routine (see Appendix) have been subtracted from the data points.

Fig. 4—(*a*) Internally illuminated polarization arising from a thick disk (opening angle 33°) as a function of albedo for various inclination angles. The equatorial optical depth of the disk is $\tau_{eq} = 1$. Dashed lines represent polarization arising from singly scattered photons, while solid lines represent the polarization arising from all scatterings. Errors are smaller than 0.1%. (*b*) Internally illuminated polarization arising from a thick disk as a function of inclination for various albedos. Line types and errors are the same as in (*a*).

Fig. 5—(*a*) As in Fig. 4*a*, but for an equatorial optical depth of $\tau_{eq} = 5$. Dotted lines indicate zero polarization. (*b*) As in Fig. 4*b*, but for an equatorial optical depth of $\tau_{eq} = 5$. Dotted lines indicate zero polarization.

Fig. 6—Typical polarization curves of the self-illuminated disk (*a*) and the externally illuminated disk (*b*). Key characteristics of these curves are labeled: $q_{DC}$ is the average base level of the disk polarization above zero, while $p_{bump}$ measures the size of the increase in total polarization at phases 0.3 and 0.7, produced by 90° scattering of photons from the external star off the disk edge. Error bars represent Stokes errors.

Fig. 7—(*a*) Self- and externally illuminated polarization arising from a thin disk (opening angle 3°) as a function of albedo for various inclination angles and two values of equatorial optical depth. $q_{DC}$ (*thick lines*) and $p_{bump}$ (*thin lines*) are defined in §3.2. Dashed lines represent polarization arising from singly scattered photons, while solid lines represent the polarization arising from all scatterings. Errors are smaller than 0.01% in $q_{DC}$ and 0.003% in $p_{bump}$. (*b*) As in Fig. 7*a*, but for a thick disk (opening angle 33°). Dotted lines indicate zero polarization; error bars are shown for Stokes errors larger than 0.02%. These figures are available in color in the electronic edition.

Fig. 8—(*a*) Self- and externally illuminated polarization arising from a thin disk (opening angle 3°) as a function of equatorial optical depth for various inclination angles and two values of albedo. Errors are smaller than 0.03% in $q_{DC}$ and 0.003% in $p_{bump}$. Line types are the same as in Fig. 7*a*. (*b*) As in Fig. 8*a*, but for a thick disk (opening angle 33°). Dotted lines indicate zero polarization; error bars are shown for Stokes errors larger than 0.005% for *a*=0.1 and 0.01% for *a*=0.9. These figures are available in color in the electronic edition.

Fig. 9—(*a*) Self- and externally illuminated polarization arising from a thin disk (opening angle 3°) as a function of $\sin^2 i$ for three values of equatorial optical depth and two values of albedo. Errors are smaller than 0.03% in $q_{DC}$ and 0.01% in $p_{bump}$. Line types are the same as in Fig. 7*a*. (*b*) As in Fig. 9*a*, but for a thick disk (opening angle 33°). Dotted lines indicate zero polarization; error bars are shown for Stokes errors larger than 0.005% for *a*=0.1 and 0.02% for *a*=0.9. These figures are available in color in the electronic edition.

Fig. 10—Variation of the total percent polarization (*p*) of an externally illuminated disk with binary phase and inclination angle for two representative cases: (*a*) a thin disk with an albedo of 0.1 and an equatorial optical depth of 10; and (*b*) a thick disk with an albedo of 0.9 and an equatorial optical depth of 1. Horizontal lines indicate zero polarization. At low inclinations, the external star contributes a significant constant polarization level, while the characteristic "bumps" near phases 0.3 and 0.7 decrease in size.

Fig. 11—(*a*) Ratio of externally illuminated polarization ($p_{bump}$) to self-illuminated polarization ($q_{DC}$) arising from a thin disk (opening angle 3°) as a function of $\sin^2 i$ for three values of equatorial optical depth and two values of albedo. Dashed lines represent polarization arising from singly scattered photons, while solid lines represent the polarization arising from all scatterings. Errors are smaller than 0.03%. We assume the two illumination sources are equally bright. (*b*) As in Fig. 11*a*, but for a thick disk (opening angle 33°); error bars are shown for Stokes errors larger than 0.25%. We have suppressed $\sin^2 i = 1$ because of large fluctuations in $p_{bump}/q_{DC}$ and correspondingly large error bars in this region.

Fig. 12—(*a*) As in Fig. 11*a*, but for the ratio of self-illuminated polarization ($q_{DC}$) to internally illuminated polarization ($q_{in}$). (*b*) As in Fig. 12*a*, but for a thick disk (opening angle 33°). Error bars are shown for Stokes errors larger than 0.04%.

Captions for color figures:

Fig. 7—(*a*) Polarization arising from a thin disk (opening angle 3°) as a function of albedo for various inclination angles and two values of equatorial optical depth. $q_{DC}$ (*blue lines*) and $p_{bump}$ (*red lines*) are defined in §3.3.1. Dashed lines represent polarization arising from singly scattered photons, while solid lines represent the polarization arising from all scatterings. Errors are smaller than 0.01% in $q_{DC}$ and 0.003% in $p_{bump}$. (*b*) As in Fig. 7*a*, but for a thick disk (opening angle 33°). Dotted lines indicate zero polarization; error bars are shown for Stokes errors larger than 0.02%.

Fig. 8—(*a*) Polarization arising from a thin disk (opening angle 3°) as a function of equatorial optical depth for various inclination angles and two values of albedo. Errors are smaller than 0.03% in $q_{DC}$ and 0.003% in $p_{bump}$. Line types are the same as in Fig. 7*a*. (*b*) As in Fig. 8*a*, but for a thick disk (opening angle 33°). Dotted lines indicate zero polarization; error bars are shown for Stokes errors larger than 0.005% for *a*=0.1 and 0.01% for *a*=0.9.

Fig. 9—(*a*) Polarization arising from a thin disk (opening angle 3°) as a function of $\sin^2 i$ for three values of equatorial optical depth and two values of albedo. Errors are smaller than 0.03% in $q_{DC}$ and 0.01% in $p_{bump}$. Line types are the same as in Fig. 7*a*. (*b*) As in Fig. 9*a*, but for a thick disk (opening angle 33°). Dotted lines indicate zero polarization; error bars are shown for Stokes errors larger than 0.005% for *a*=0.1 and 0.02% for *a*=0.9.